\documentclass[prc,twocolumn,twoside,showpacs,nofootinbib,floatfix]{revtex4}

\usepackage{graphicx,color,rotating}
\usepackage{amsmath,amssymb,bm}
\usepackage{ae}

\newcommand{\ket}[1]{\ensuremath{\,|{#1}\rangle}}

\newcommand{\matrixe}[3]{\ensuremath{\langle{#1}|\,{#2}\,|{#3}\rangle}}

\newcommand{\comm}[2]{\ensuremath{[{#1},{#2}]}}

\newcommand{\op}[1]{\ensuremath{#1}}
\newcommand{\adj}[1]{\ensuremath{{{#1}}^{\dag}}}
\newcommand{\corr}[1]{\ensuremath{\widetilde{#1}}}

\renewcommand{\vec}[1]{\ensuremath{\bm{#1}}}

\newcommand{\gO}{\ensuremath{\op{g}}}

\newcommand{\pO}{\ensuremath{\op{p}}}
\newcommand{\qO}{\ensuremath{\op{q}}}
\newcommand{\rO}{\ensuremath{\op{r}}}

\newcommand{\vO}{\ensuremath{\op{v}}}

\newcommand{\etaO}{\ensuremath{\op{\eta}}}

\newcommand{\CO}{\ensuremath{\op{C}}}

\newcommand{\HO}{\ensuremath{\op{H}}}

\newcommand{\OO}{\ensuremath{\op{O}}}

\newcommand{\TO}{\ensuremath{\op{T}}}
\newcommand{\UO}{\ensuremath{\op{U}}}
\newcommand{\VO}{\ensuremath{\op{V}}}

\newcommand{\CCO}{\ensuremath{\adj{\op{C}}}}
\newcommand{\UUO}{\ensuremath{\adj{\op{U}}}}

\newcommand{\qOV}{\ensuremath{\vec{\op{q}}}}
\newcommand{\rOV}{\ensuremath{\vec{\op{r}}}}

\newcommand{\LOV}{\ensuremath{\vec{\op{L}}}}

\newcommand{\sigmaOV}{\ensuremath{\vec{\op{\sigma}}}}

\newcommand{\Tint}{\ensuremath{\TO_\text{int}}}

\newcommand{\tensorRRO}{\ensuremath{\op{S}_{12}(\tfrac{\rOV}{\rO},\tfrac{\rOV}{\rO})}}

\newcommand{\tensorRQO}{\ensuremath{\op{S}_{12}(\rOV,\qOV_{\Omega})}}

\newcommand{\spinorbitO}{\ensuremath{(\vec{\op{L}}\cdot\vec{\op{S}})}}

\newcommand{\Rp}{\ensuremath{R_+}}

\newcommand{\UCOM}{\ensuremath{\textrm{UCOM}}}

\newcommand{\fm}{\ensuremath{\,\text{fm}}}

\definecolor{FGViolet}{rgb}{0.61,0.32,0.61}
\definecolor{FGDarkBlue}{rgb}{0,0,0.6}
\definecolor{FGBlue}{rgb}{0,0,0.8}
\definecolor{FGLightBlue}{rgb}{0.2, 0.6, 0.8}
\definecolor{FGGreen}{rgb}{0.2,0.7,0.2}
\definecolor{FGLightGreen}{rgb}{0.4,1,0.4}
\definecolor{FGYellow}{rgb}{1,0.95,0}
\definecolor{FGOrange}{rgb}{0.95,0.5,0.1}
\definecolor{FGRed}{rgb}{0.8,0,0}
\definecolor{FGWhite}{rgb}{1,1,1}
\definecolor{FGLightGray}{rgb}{0.8,0.8,0.8}
\definecolor{FGGray}{rgb}{0.5,0.5,0.5}
\definecolor{FGDarkGray}{rgb}{0.3,0.3,0.3}
\definecolor{FGBlack}{rgb}{0,0,0}

\newcommand{\linethinsolid}[1][black]{\unitlength 1ex
  {\color{#1}
  \begin{picture}(6,1)
  \linethickness{0.12mm}
  \put(0,0.5){\line(1,0){6.0}}
  \end{picture}}\nolinebreak
}

\newcommand{\linethindashed}[1][black]{\unitlength 1ex
  {\color{#1}
  \begin{picture}(6,1)
  \linethickness{0.12mm}
  \put(0,0.5){\line(1,0){1.5}}
  \put(2.2,0.5){\line(1,0){1.5}}
  \put(4.4,0.5){\line(1,0){1.5}}
  \end{picture}}\nolinebreak
}

\newcommand{\linethindotted}[1][black]{\unitlength 1ex
  {\color{#1}
  \begin{picture}(6,1)
  \linethickness{0.12mm}
  \put(0,0.5){\line(1,0){0.8}}
  \put(1.2,0.5){\line(1,0){0.8}}
  \put(2.4,0.5){\line(1,0){0.8}}
  \put(3.6,0.5){\line(1,0){0.8}}
  \put(4.8,0.5){\line(1,0){0.8}}
  \end{picture}}\nolinebreak
}

\newcommand{\linethindashdot}[1][black]{\unitlength 1ex
  {\color{#1}
  \begin{picture}(6,1)
  \linethickness{0.12mm}
  \put(0,0.5){\line(1,0){0.4}}
  \put(0.9,0.5){\line(1,0){1.5}}
  \put(2.9,0.5){\line(1,0){0.4}}
  \put(3.8,0.5){\line(1,0){1.5}}
  \put(5.8,0.5){\line(1,0){0.4}}
  \end{picture}}\nolinebreak
}

\begin{document}

\title{The Unitary Correlation Operator Method from a \\
       Similarity Renormalization Group Perspective}

\author{H. Hergert}
\email{Heiko.Hergert@physik.tu-darmstadt.de}

\author{R. Roth}
\email{Robert.Roth@physik.tu-darmstadt.de}

\affiliation{Institut f\"ur Kernphysik, Technische Universit\"at Darmstadt,
64289 Darmstadt, Germany}

\date{\today}

\begin{abstract}    
We investigate how the Unitary Correlation Operator Method (UCOM), developed to explicitly describe the strong short-range central and tensor correlations present in the nuclear many-body system, relates to the Similarity Renormalization Group (SRG), a method to band-diagonalize Hamiltonians by continuous unitary transformations. We demonstrate that the structure of the UCOM transformation, originally motivated from the physically intuitive picture of correlations in coordinate space, arises naturally from the SRG flow equation. Apart from formal considerations we show that the momentum space matrix elements of the effective interactions obtained in both schemes agree extremely well.
\end{abstract}


\maketitle

\clearpage

The key ingredient in modern \emph{ab initio} studies of nuclear structure are realistic nucleon-nucleon interactions, which reproduce the experimental two-nucleon observables such as scattering phase-shifts with high precision. Different realistic interactions have been constructed using phenomenological formulations (Argonne V18),  the meson-exchange picture (CD Bonn), or systematic expansions within chiral effective field theory (chiral N3LO) which also yield a consistent description of multi-nucleon forces
(for reviews, see \cite{Machleidt:2001rw, Epelbaum:2005pn}). Generally, when used, e.g., in a no-core shell model framework \cite{Navratil:2000gs}, these interactions exhibit very slow convergence with increasing model-space size \cite{Roth:2005pd}. The reason are large off-diagonal matrix elements which connect the low-energy and low-momentum states of the model space with high-lying states typically not included in the model space. In other words, simple model spaces do not allow for a description of the correlations that are induced by realistic NN interactions.
  
In order to address this issue, different schemes have been proposed to derive soft effective interactions that preserve the physically constrained properties of the original potential, the phase-shifts.
One of those, the renormalization group approach \cite{Bogner:2003wn}, introduces a momentum cutoff and  explicitly decouples low and high-momentum components. Eventually one obtains a low-momentum interaction $V_{\text{low-}k}$ which is only defined below the cutoff scale.

Another approach, the Unitary Correlation Operator Method (UCOM) \cite{Feldmeier:1997zh, Neff:2002nu}, was devised as a means to describe the dominant correlations induced by the short-range repulsion and the tensor interaction by an explicit unitary transformation. The unitary transformation of the Hamiltonian leads to a phase-shift equivalent correlated interaction $\VO_\UCOM$ which is suitable for simple model spaces. It has been successfully employed in a wide range of many-body calculations, from the no-core shell model to Hartree-Fock and its extensions \cite{Roth:2004ua, Roth:2005pd, Roth:2005ah, Paar:2006ua, Papakonstantinou:2006vc}. In these calculations, one benefits from the dramatically improved convergence behavior of the correlated interaction, which results from a pre-diagonalization of the interaction in momentum space. The observed band-diagonal structure indicates that low and high momenta have been decoupled, but in a different manner than in the $V_{\text{low-}k}$ or chiral EFT approach, which give just the low-momentum block of a block-diagonal Hamiltonian.

Recently, Bogner \emph{et al.} \cite{Bogner:2006pc, Bogner:2007jb} have proposed the application of the Similarity Renormalization Group (SRG) to the NN interaction, and demonstrated the derivation of an effective interaction in this framework. The SRG approach, although starting from a different conceptional background, has formal similarities to the UCOM approach. In this paper we discuss how the UCOM and the SRG approach can be formally related, how the resulting effective interactions compare, and what can be learned for the further development of both schemes. 

The formal structure of the UCOM is motivated by physical considerations on the structure and origin of the dominant many-body correlations. The short-range repulsion in the central part of the NN interaction drives the interacting nucleon pair apart. The tensor interaction induces correlations between the relative distance and the spin of the nucleon pair, leading to the characteristic mixing between components with relative orbital angular momentum $L$ and $L\pm2$ in the $S=1$ channel. To imprint these correlations on a many-body state, we construct a unitary transformation with the generators
\begin{equation}\label{eq_gen_r}
  \gO_r = \frac{1}{2}\left(\qO_r s(r) + s(r)\qO_r\right)
\end{equation}
and
\begin{align}\label{eq_gen_tens}
  \gO_\Omega &= \vartheta(r)\tensorRQO \notag \\
   &\equiv\vartheta(r)\frac{3}{2}
         \left(\left(\sigmaOV_1\!\cdot\!\qOV_{\Omega}\right)\left(\sigmaOV_2\!\cdot\!\rOV\right)+\right.         
         \left.\left(\sigmaOV_1\!\cdot\!\rOV\right)\left(\sigmaOV_2\!\cdot\!\qOV_{\Omega}\right)\right)\,,
\end{align}
where 
\begin{gather}
  \qO_r \equiv \frac{1}{2}\left(\qOV\cdot\frac{\rOV}{r} + \frac{\rOV}{r}\cdot\qOV\right)\,,\\
  \qOV_\Omega \equiv \qOV-\qO_r\frac{\rOV}{r}=\frac{1}{2r^2}\left(\LOV\times\rOV-\rOV\times\LOV\right)\,.
\end{gather}
The generator $\gO_r$ uses the radial part of the relative momentum operator $\qOV$ to create a shift in the radial direction, while $\gO_\Omega$ is constructed from the so-called orbital momentum, i.e. the angular part of $\qOV$, and generates shifts perpendicular to $\rOV$. The strength and range of the transformation is governed by the central and tensor correlation functions $s(r)$ and $\vartheta(r)$, respectively, which are determined separately in each $(S,T)$-channel by means of an energy minimization \cite{Neff:2002nu, Roth:2004ua}. The unitary transformation is then written as 
\begin{equation}\label{eq_simtrans}
  \CO\equiv\CO_\Omega\CO_r\equiv\exp\bigg(\!-i\sum_{j<k}\gO_{r,jk}\bigg)\!\exp\bigg(\!-i\sum_{j<k}\gO_{\Omega,jk}\bigg)\,,
\end{equation}
where the sum runs over all nucleon pairs. One can now proceed to calculate expectation values either by applying $\CO$ to the many-body state $\ket{\Psi}$ or to a given observable $\OO$, yielding either a correlated state $\ket{\corr{\Psi}}$ or a correlated operator $\corr{\OO}$:
\begin{equation}
  \matrixe{\corr{\Psi}}{\OO}{\corr{\Phi}}=\matrixe{\Psi}{\CCO_r\CCO_\Omega\OO\CO_\Omega\CO_r}{\Phi}=
  \matrixe{\Psi}{\corr{\OO}}{\Phi}\,.
\end{equation}
Using correlated operators is more convenient in applications, but one has to be aware that all observables of interest need to be correlated consistently in that case. 

From the structure of the transformation \eqref{eq_simtrans}, it is evident that a correlated operator $\corr{\OO}$ will always be of $A$-body type in Fock space, where $A$ is the number of nucleons. The decomposition into  irreducible contributions $\corr{\OO}^{[n]}$ for a specific particle number $n\leq A$ leads to the following cluster expansion for the correlated operator:
\begin{equation}
  \corr{\OO}=\CCO\OO\CO=\corr{\OO}^{[1]}+\corr{\OO}^{[2]}+\ldots+\corr{\OO}^{[A]}\,.
\end{equation}
If the range of the correlation functions is small compared to the mean inter-particle distance, we can employ the two-body approximation and omit negligible cluster terms beyond the second order (for details see Refs. \cite{Feldmeier:1997zh, Neff:2002nu, Roth:2005pd}). For the construction of the correlated Hamiltonian in two-body approximation, it is then sufficient to consider the Hamiltonian in the two-nucleon system, 
\begin{equation}\label{eq_hrel}
  \HO = \Tint+\VO \equiv \frac{\qOV^2}{2\mu} + \VO\,,
\end{equation}
where we have already subtracted the center-of-mass kinetic energy, which is not affected by the correlation procedure. Applying the correlation operators, 
\begin{equation}
  \CCO_r\CCO_\Omega\HO\CO_\Omega\CO_r = \TO_\text{int} + \corr{\TO}_\text{int}^{[2]}+\corr{\VO}^{[2]}+\ldots\,,
\end{equation}
and collecting the two-body contributions from the correlated kinetic energy and the transformed interaction, we obtain the effective interaction $\VO_{\UCOM}$: 
\begin{equation}
  \VO_\UCOM\equiv\corr{\TO}_\text{int}^{[2]}+\corr{\VO}^{[2]}\,.
\end{equation}
Based on this definition, we can construct an explicit operator representation of $\VO_\UCOM$, which illustrates how the transformation affects the structure of the interaction \cite{Roth:2005pd}. The central correlator merely causes a transformation of the relative coordinate of the nucleon pair, 
\begin{equation}\label{eq_corr_r}
  \CCO_r\rO\CO_r = \Rp(\rO)\,,
\end{equation}
where $\Rp(r)$ is related to $s(r)$ by the integral equation
\begin{equation}
  \int_{r}^{\Rp(r)}\frac{d\xi}{s(\xi)}=1\,.
\end{equation}
The action of the tensor correlator needs to be evaluated via a Baker-Campbell-Hausdorff (BCH) series, which does not terminate in general. As a result of the $\CO_\Omega$-transformation, the interaction operator acquires new momentum-dependent tensor terms \cite{Roth:2004ua}, which are not present in typical parametrizations of realistic interactions. 

The matrix elements of $\VO_\UCOM$ in a partial-wave basis can be evaluated directly by applying the tensor correlation operators to the two-body states, avoiding the complications of the BCH expansion \cite{Roth:2005pd}. The correlated matrix elements in an harmonic oscillator basis have been used for no-core shell model calculations, which demonstrate the rapid convergence which results from the band-diagonal structure of the correlated matrix elements \cite{Roth:2005pd}. 

This provides a first heuristic link to the Similarity Renormalization Group (SRG) approach. Following Wegner's formulation of the SRG \cite{Wegner:1994}, one can write down a flow equation for the many-body Hamiltonian $\HO$. Denoting the flow parameter $\alpha$, the operator evolves via 
\begin{equation} \label{eq_h_flow}
  \frac{d\HO_\alpha}{d\alpha}=\comm{\etaO(\alpha)}{\HO_\alpha}\,,\quad\HO_0=\HO\,,
\end{equation}
with
\begin{equation}
  \HO_\alpha\equiv \UO(\alpha)\HO\UUO(\alpha)\equiv\TO_\text{int}+\VO_\alpha\,,
\end{equation}
where all $\alpha$-dependent contributions have been absorbed in the many-body interaction $\VO_\alpha = \HO_\alpha-\TO_\text{int}$ and the $\alpha$-independent intrinsic kinetic energy $\TO_\text{int}=\TO-\TO_{\text{cm}}$ has been separated. The anti-hermitian generator $\etaO(\alpha)$ formally satisfies 
\begin{equation}
  \etaO(\alpha)=\frac{d\UO(\alpha)}{d\alpha}\UUO(\alpha)=-\adj{\etaO}(\alpha)\,,
\end{equation}
and has to be chosen appropriately for practical applications. Wegner's original choice was
\begin{equation}
  \etaO(\alpha)=\comm{\text{diag}(\HO_\alpha)}{\HO_\alpha}\,,
\end{equation}
which one can understand intuitively: if $\HO_\alpha$ commutes with its diagonal part with respect to a certain basis, then the generator vanishes and one has reached a fixed point of the flow. Trivial cases aside, this can only happen if $\HO_\alpha$ is, in fact, diagonal in that basis. Thus, the generator \emph{dynamically} drives $\HO_\alpha$ towards a diagonal structure with increasing $\alpha$. The price one has to pay for this simplification is that one has to deal with complicated many-body interactions in $\VO_\alpha$ even if one starts with a two-body potential \cite{Wegner:2000gi}. 

A simpler choice for the generator was suggested by Szpigel and Perry \cite{Szpigel:2000xj} and employed successfully by Bogner et al. \cite{Bogner:2006pc}. First of all, one confines the evolution to a two-body space, discarding induced multi-nucleon interactions from the outset --- this assumption corresponds to the two-body approximation used in the UCOM framework. The two-body generator is given by  
\begin{equation}\label{eq_gen}
  \etaO(\alpha)=\comm{\Tint}{\HO_\alpha}=\comm{\frac{\qOV^2}{2\mu}}{\HO_\alpha}\,,
\end{equation}
which aims to diagonalize the two-body Hamiltonian $\HO_\alpha$ in a basis of eigenstates of both $\pO_r^2$ and $\LOV^2$. Hence, in a partial-wave momentum-space basis $\ket{q\,(LS)J\,T}$ this generator drives the matrix elements towards a band-diagonal structure with respect to $(q,q')$ and $(L,L')$.  

Let us consider the Argonne V18 \cite{Wiringa:1994wb} potential as a specific example of a realistic NN interaction. Its operator structure in a given spin-isospin $(S,T)$ channel can be written as
\begin{equation}\label{eq_v_general}
  \VO_{ST}=\sum_p v^{ST}_p(\rO)\,\OO_p\,,
\end{equation}
where $\OO_p \in \left\{1, \LOV^2, \spinorbitO, \spinorbitO^2, \tensorRRO \right\}$. The charge-dependent terms are neglected, as they are of no immediate concern for the following discussion.

Using this characteristic operator form of the interaction we can evaluate the generator \eqref{eq_gen} at $\alpha=0$, which determines the initial flow in \eqref{eq_h_flow}:
\begin{equation}\label{eq_gen0}
  \etaO(0)=\comm{\Tint}{\HO_0}=\frac{1}{2\mu}\comm{\qO_r^2+\frac{\LOV^2}{\rO^2}}{\VO}\,.
\end{equation}
For the commutator with $\qO_r^2$ we find in each $(S,T)$-channel (omitting the $ST$-index for brevity),
\begin{equation}
\begin{split}
  \comm{\qO_r^2}{\VO}
  &= \sum_p \comm{\qO_r^2}{\vO_p(\rO)O_p} \\
  &= -i \sum_p \left(\qO_rv'_p(\rO)+v'_p(\rO)\qO_r\right)\OO_p\,,
\end{split}
\end{equation}
since  $q_r$ commutes with all of the operators $\OO_p$. For the angular component we get
\begin{equation}
\begin{split}
  \comm{\frac{\LOV^2}{\rO^2}}{\VO} 
  &= \comm{\frac{\LOV^2}{\rO^2}}{v_t(\rO)\tensorRRO} \\
  &= -4i\frac{v_t(\rO)}{\rO^2}\tensorRQO\,,
\end{split}
\end{equation}
because all but the tensor component of the interaction commute with $\LOV^2$. Thus, we obtain the following form for the initial generator 
\begin{equation}
  \etaO(0) =  
  \frac{i}{2}(\qO_r S(\rO) + S(\rO)\qO_r) + i \Theta(\rO)\tensorRQO \,,
\end{equation}
with
\begin{equation}\label{eq_corrfunc_srg}
  S(\rO) \equiv -\frac{1}{\mu} \bigg(\sum_pv'_p(\rO)\OO_p \bigg)\quad\text{and}\quad
  \Theta(\rO) \equiv -\frac{2}{\mu}\frac{v_t(\rO)}{\rO^2}\,. 
\end{equation}
This is a remarkable result: the structure of $\etaO(0)$ resembles the sum of the UCOM generators $\gO_r$ and $\gO_\Omega$ given in Eqs. \eqref{eq_gen_r} and \eqref{eq_gen_tens}. The symmetrized radial momentum operator and the non-trivial momentum-dependent tensor operator $\tensorRQO$, which have been constructed based on the structure of the short-range correlations in the UCOM framework, result directly from the commutation relation defining the generator of renormalization group flow. This result connects the SRG flow picture with a physically intuitive picture of central and tensor correlations in a many-body state. 

Conversely, the flow equation confirms that the UCOM ansatz contains the important generators. Moreover, it provides guidance for generalizing the UCOM scheme. So far, the correlation function $s(r)$ in Eq. \eqref{eq_gen_r} describing the central correlations in the UCOM framework only included a  $(S,T)$-dependence. However, in the SRG framework the corresponding $S(r)$ in \eqref{eq_corrfunc_srg} is an operator-valued function which is sensitive to orbital and total angular momentum as well. This amounts to a simple extension of the UCOM scheme using different correlation functions $s(r)$ for each partial-wave.

Let us return to the flow equation \eqref{eq_h_flow} to discuss one inherent difference between the SRG and the UCOM scheme. By solving Eq. \eqref{eq_h_flow}, we evolve $\HO_\alpha$ along a non-linear trajectory in the manifold of unitarily equivalent operators towards the fixed point, i.e. (band-) diagonality. The structure of the generator $\etaO(\alpha)$ will adapt \emph{dynamically} at each step of the flow, starting from $\etaO(0)$. The UCOM scheme, in contrast, uses a static generator to evolve the Hamiltonian in a single step from the initial point to the final point of the flow trajectory. If one would assume an SRG generator independent of $\alpha$, the flow equation could be integrated, 
\begin{equation}
  \HO_\alpha=
     e^{-\alpha\etaO(0)} \HO_0 e^{\alpha\etaO(0)}\,,
\end{equation}
and one would recover an explicit unitary transformation as in the UCOM case, corresponding to a linear flow trajectory along the direction specified by $\etaO(0)$. A further, merely technical difference is that the UCOM formulation assumes separate unitary transformations \eqref{eq_simtrans} with $\gO_r$ and $\gO_\Omega$. In the SRG picture this amounts to separate evolutions with two flow parameters $\alpha_r$ and $\alpha_\Omega$ using the generators
\begin{equation}
  \etaO_r=\frac{1}{2\mu}\comm{\qO_r^2}{\HO_{\alpha_r}} 
  \quad\text{and}\quad
  \etaO_\Omega=\frac{1}{2\mu}\comm{\frac{\LOV^2}{\rO^2}}{\HO_{\alpha_\Omega}}\,,
\end{equation}
respectively. 

The difference between the static generators of the UCOM and the dynamic generators of the SRG affect the choice of the correlation functions. One could be tempted to use the radial dependencies of the initial SRG generator \eqref{eq_corrfunc_srg} in UCOM, since they are directly given by the original potential. However, this is not appropriate since the UCOM correlation functions effectively integrate over the complete flow trajectory. One would have to use the dynamic flow picture to determine the UCOM correlation functions that directly connect initial and final point of the flow trajectory. This is an exciting prospect and subject of further investigations.

The current approach for constructing the UCOM correlation functions employs an energy minimization in the two-nucleon system, considering the lowest partial waves for each $(S,T)$-channel \cite{Roth:2005pd}. Let us consider the $^3S_1$ partial wave with momentum eigenstates $\ket{q(01)10}$. Assuming vanishing relative momentum $q$, i.e. a constant radial wavefunction, we minimize the expectation value of the correlated two-body Hamiltonian 
\begin{equation}
  E_{10}=\matrixe{0(01)10}{\CCO_r\CCO_\Omega\HO\CO_\Omega\CO_r}{0(01)10}
\end{equation}
by varying the parameters of suitable parametrizations of the central and tensor correlation functions (see Ref. \cite{Roth:2005pd} for details). The range of the tensor correlation function is constrained using the value of the integral 
\begin{equation}\label{eq_constraint}
  I_\vartheta = \int dr r^2\vartheta(r)\,.
\end{equation}
The value of $I_\vartheta$ in the $(S,T)=(1,0)$ channel constitutes the main parameter of $\VO_\UCOM$, which can be understood by scale arguments in coordinate space: the repulsive core has a very steep slope at about $0.5\,\fm$, and therefore represents a distinct and physically motivated length scale. The case is different for the tensor interaction, which is of considerable strength up to $2-3\,\fm$. By restricting the range of the tensor correlator, one introduces a scale separation --- correlations at shorter length scales are described by the correlation operator, while long-range correlations need to be described by the Hilbert space \cite{Roth:2005pd}.

\begin{figure}[t]
\includegraphics[width=0.95\columnwidth]{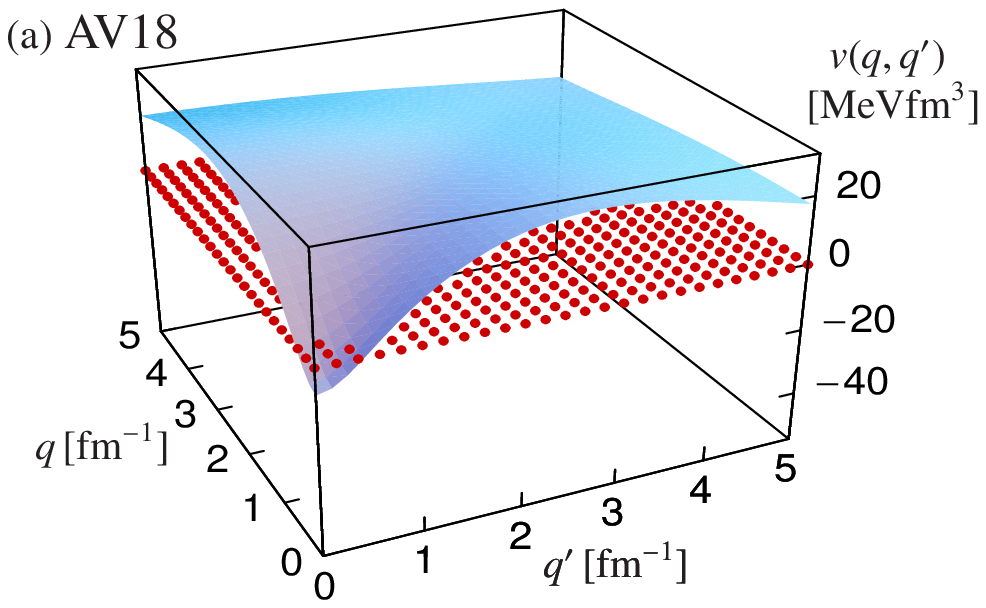}\\[4pt]
\includegraphics[width=0.95\columnwidth]{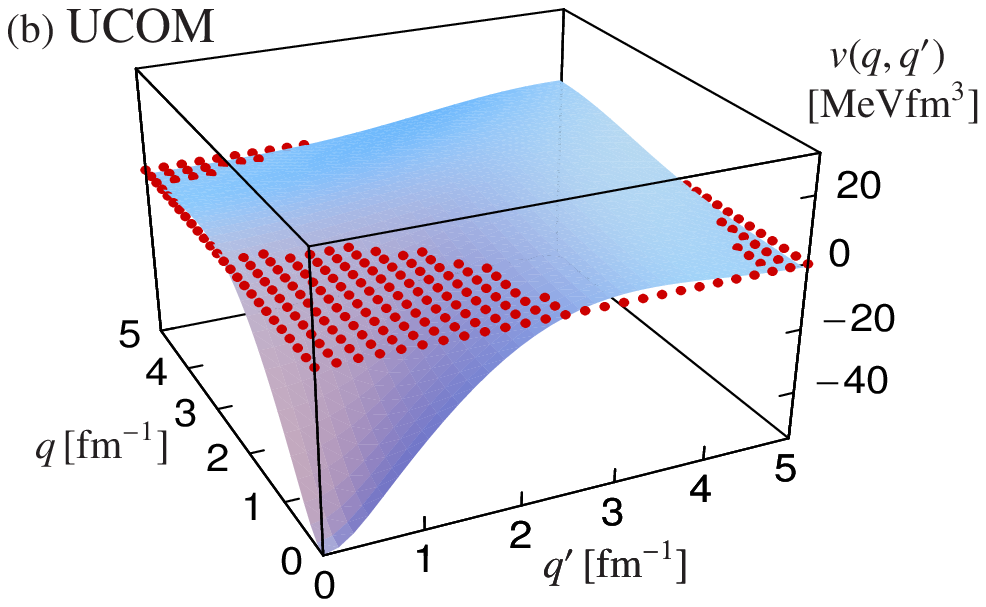}\\[4pt]
\includegraphics[width=0.95\columnwidth]{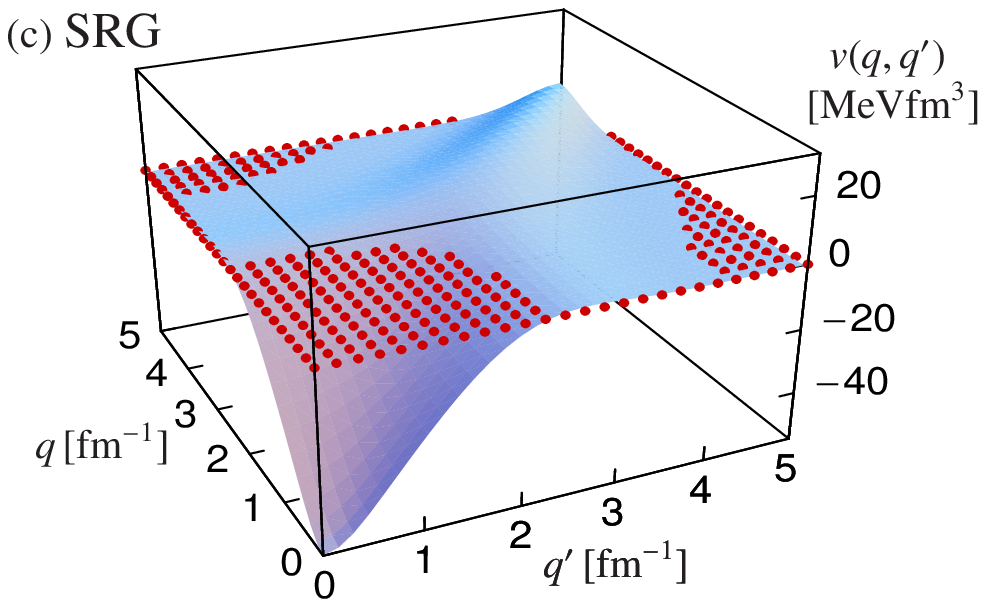}
\caption{(color online) Three-dimensional plots of the momentum space matrix elements $v(q,q')$ in the $^3S_1$ partial wave for the AV18 potential (a), the $\VO_\UCOM$ interaction with $I_\vartheta=0.09\fm^3$ (b), and the SRG interaction $\VO_\alpha$ for $\bar{\alpha}=0.0215\fm^4$ (c). The dots mark the plane of vanishing matrix elements.}
\label{fig_me_3d}
\end{figure}

That the constrained energy minimization has the desired effect becomes evident if we compare the matrix elements of $\VO_\UCOM$ with those of $\VO_\alpha$ evolved by solving the full flow equation \eqref{eq_h_flow} in momentum space, as described in Ref. \cite{Bogner:2006pc}. In Fig. \ref{fig_me_3d}, we display the $^3S_1$ matrix elements of AV18 along with $\VO_\UCOM$ for $I_\vartheta=0.09\fm^3$ (optimized for calculations without $3N$-forces \cite{Roth:2005pd}), and $\VO_\alpha$ for $\bar{\alpha}=\frac{\alpha}{2\mu}=0.0215\fm^4$. This value of $\bar{\alpha}$ was fixed by matching the matrix elements of $\VO_\UCOM$ and $\VO_\alpha$ at $q=q'=0\fm^{-1}$. 

While the matrix elements of the parent interaction have large off-diagonal contributions and strongly repulsive diagonal terms, both $\VO_\UCOM$ and $\VO_\alpha$ exhibit a dramatic suppression of the off-diagonal terms and an enhancement of the attractive low-momentum component. The $\VO_\UCOM$ and $\VO_\alpha$ matrix elements are very similar, the only noticeable difference being the stronger increase of the SRG matrix elements at high momenta when approaching the diagonal.

\begin{figure}
\includegraphics[width=\columnwidth]{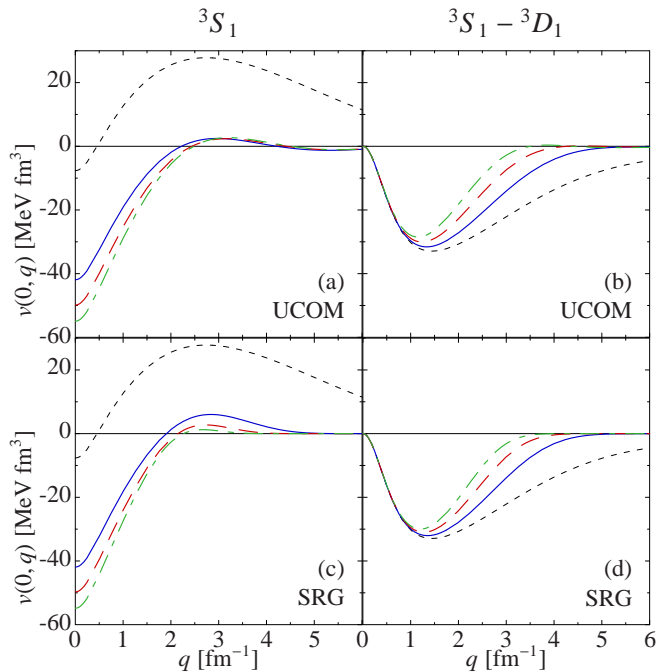}
\caption{(color online) Off-diagonal matrix elements $v(0,q)$ in the $^3S_1$ and $^3S_1-^3D_1$ partial waves. Panels (a) and (b) show the $\VO_\UCOM$ matrix elements for $I_\vartheta=$ 
$0.03 \fm^3$ (\linethinsolid[FGBlue]), $0.06 \fm^3$ (\linethindashed[FGRed]), and $0.09 \fm^3$ (\linethindashdot[FGGreen]) in comparison to those of AV18 (\linethindotted[FGGray]). Panels (c) and (d) show the SRG-evolved matrix elements of $\VO_\alpha$ for flow parameter values $\bar{\alpha}=0.0055 \fm^4$ (\linethinsolid[FGBlue]), $0.0125 \fm^4$ (\linethindashed[FGRed]), and $0.0215 \fm^4$ (\linethindashdot[FGGreen]), which have been chosen to reproduce the value of the corresponding $\VO_\UCOM$ matrix element at $q=q'=0\fm^{-1}$.}
\label{fig_me_offdiag}
\end{figure}

As discussed above, the range constraint $I_\vartheta$ is the main free parameter of $\VO_\UCOM$, similar to $\alpha$ for the SRG evolution --- although this analogy should not be taken too far. In Fig. \ref{fig_me_offdiag}, we compare the off-diagonal matrix elements $\vO(0,q)$ in the $^3S_1$ and $^3S_1-^3D_1$ partial waves for various values of $I_\vartheta$ and $\alpha$, respectively. We find that with increasing $I_\vartheta$ the decoupling of different momenta is improved, much like the increase of $\alpha$ renders $\VO_\alpha$ more diagonal. In the 
$^3S_1$ partial wave, the SRG evolution leads to vanishing off-diagonal matrix elements at large $q$. In comparison, the matrix elements of $\VO_\UCOM$ reveal tiny residual contributions at large momenta, which are most likely related to the particular parametrization used for the UCOM correlation functions and can probably be remedied by using refined parametrizations or extracting them directly from solutions of the SRG flow equation. Given that the correlation functions were optimized using the diagonal $^3S_1$ matrix element at $q=q'=0\fm^{-1}$ only, the agreement between $\VO_\UCOM$ and $\VO_\alpha$ is impressive. The same observation holds for the coupled channel $^3S_1-^3D_1$, which was not considered in the optimization of the UCOM correlators at all: the effects of the UCOM and the SRG evolution on the original interaction are almost indistinguishable. Both schemes have the same efficiency in suppressing the off-diagonal contributions in this tensor-dominated partial wave.

In summary, we have shown that the UCOM and the SRG, although approaching the construction of effective nucleon-nucleon intercations from different conceptional backgrounds --- correlations in a coordinate space picture for UCOM, diagonality in momentum space for the SRG --- are closely related. Each of the methods provide insight into the other: from the UCOM, we gain a physically intuitive picture for the actual effects of the more abstract SRG flow equation. The SRG approach, on the other hand, suggests ways to construct the correlation functions in a systematic fashion, without resorting to parametrizations or using any parameters aside from the scale $\alpha$, and immediately yields the structure of the UCOM generators. This will prove helpful when correlations in the $3N$ system are studied, because the choice of a \emph{genuine} three-nucleon correlation operator in the UCOM is not obvious. In the three-nucleon system, the SRG flow equation will likely facilitate the treatment of three-body correlations which are induced by the unitary transformation of the interaction in both approaches, and which have been omitted thus far. The impressive agreement of both approaches in momentum space indicates that the approximations required to derive the UCOM in its present form from the SRG --- separate evolutions using the scale-independent generators $\etaO_r$ and $\etaO_\Omega$ --- are remedied by the optimization of the correlation functions with physical input. 

The construction of the UCOM correlators via the SRG approach, as well as the construction of SRG interactions in an operator representation and the explicit inclusion of three-nucleon interactions are interesting topics for further research.

\section*{Acknowledgments}

This work is supported by the Deutsche Forschungsgemeinschaft through contract SFB 634. 


\end{document}